\documentclass[twocolumn]{aastex631}
\hypersetup{colorlinks,linkcolor={cyan},citecolor={cyan},urlcolor={cyan}} 

\usepackage{lineno}

\usepackage{color}

\defcitealias{Y20}{Y20}

\def\ergs{\textit{$\rm erg\ s^{-1}$}}
\def\loglbol{\textit{${\rm log}\,L_{\rm bol}$}}
\def\lbol{\textit{$L_{\rm bol}$}}
\def\Rin{\textit{$R_{\rm in}$}}
\def\Reff{\textit{$R_{\rm eff}$}}

\def\cerr{\textit{$C_{\rm noise}$}}
\def\nerr{\textit{$N_{\rm err}$}}

\def\famp{\textit{$f_{\rm amp}$}}
\def\trans{\textit{$\phi(\tau)$}}

\def\SFobs{\textit{$\rm SF_{obs}$}}
\def\um{\textit{$\rm \mu m$}}
\def\lambdar{\textit{$\lambda_{\rm rest}$}}

\def\SFinf{\textit{$\rm SF_{\infty}$}}

\def\SFwdisk{\textit{${\rm SF}_{w, \rm AD}$}}
\def\SFwtorus{\textit{${\rm SF}_{w, \rm torus}$}}
\def\SFg{\textit{${\rm SF}_{g, \rm AD}$}}

\def\SFw{\textit{${\rm SF}_w$}}

\begin{document}

\title{Constraining AGN Torus Sizes with Optical and Mid-Infrared Ensemble Structure Functions}

\author{Junyao Li}
\affiliation{Department of Astronomy, University of Illinois at Urbana-Champaign, Urbana, IL 61801, USA}
\correspondingauthor{Junyao Li}
\email{junyaoli@illinois.edu}

\author[0000-0003-1659-7035]{Yue Shen}
\affiliation{Department of Astronomy, University of Illinois at Urbana-Champaign, Urbana, IL 61801, USA}
\affiliation{National Center for Supercomputing Applications, University of Illinois at Urbana-Champaign, Urbana, IL 61801, USA}

\begin{abstract}
We propose a new method to constrain the size of the dusty torus in broad-line active galactic nuclei (AGNs) using optical and mid-infrared (MIR) ensemble structure functions (SFs). Because of the geometric dilution of the torus, the MIR response to optical continuum variations has suppressed variability with respect to the optical that depends on the geometry (e.g., size, orientation, opening angle) of the torus. More extended tori have steeper MIR SFs with respect to the optical SFs. We demonstrate the feasibility of this SF approach using simulated AGN light curves and a geometric torus model. While it is difficult to use SFs to constrain the orientation and opening angle due to insensitivity of the SF on these parameters, the size of the torus can be well determined. Applying this method to the ensemble SFs measured for 587 SDSS quasars, we measure a torus $R-L$ relation of {${\rm log}\,\Reff\, (\rm pc) = 0.51_{-0.04}^{+0.04} \times {\rm log}\,(\lbol/10^{46}\,\ergs) -0.38_{-0.01}^{+0.01}$} {in the WISE $W1$ band, and sizes $\sim1.4$ times larger in the $W2$ band}, which are in good agreement with dust reverberation mapping measurements. Compared with the reverberation mapping technique, the SF method is much less demanding in data quality and can be applied to any optical+MIR light curves for which a lag measurement may not be possible, as long as the variability process and torus structure are stationary. While this SF method does not extract all information contained in the light curves (i.e., the transfer function), it provides an intuitive interpretation for the observed trends of AGN MIR SFs compared with optical SFs. 
\end{abstract}

\keywords{black hole physics --- galaxies: active --- quasars: general --- surveys}

\section{Introduction}

Active galactic nuclei (AGNs) are powered by accreting supermassive black holes (SMBHs) that produce enormous quantities of radiation across the entire electromagnetic wavelength range. It is widely accepted that the inner region of an AGN consists of a corona that produces compact X-ray emissions and an accretion disk that emits thermal blackbody spectrum in the UV/optical bands \citep{Shakura1973, Haardt1993}, although the exact structure and geometry of the X-ray corona and accretion disk are still debated \citep[e.g.,][]{Cackett2021}. The central SMBH is surrounded by a dusty torus that absorbs UV/optical photons and re-emits them in the infrared, producing a MIR bump in their spectral energy distribution (SED; \citealt{Elvis1994, Lyu2017}). The inclination angle and covering factor of the dusty torus are key ingredients in the AGN unification model that determine the appearance of AGNs as unobscured/Type~1s and obscured/Type~2s \citep{Antonucci1993, Netzer2015}. 

Variability is a ubiquitous feature of AGNs \citep{Ulrich1997}. It offers a unique means to investigate the inner structure of distant AGNs that cannot be spatially resolved, such as measuring the size of the accretion disk (a few light days; e.g., \citealt{Mudd2018, Yu2020}) and the dusty torus ($\sim0.01-1$~pc; e.g., \citealt{Suganuma2006, Kishimoto2007, Koshida2014, Minezaki2019, Lyu2019, Y20}) through  reverberation mapping. The variability of the continuum emission of AGNs in the optical bands can be well characterized by the damped random walk (DRW) model over months to years timescales, although its physical origin is still unclear \citep{Kelly2009, MacLeod2010, Zu2013, Burke2021,Stone2022}. The DRW model describes AGN variability as a stochastic process using an exponential covariance function 
\begin{equation}
k(\Delta t) = \sigma^2 {\rm exp}(-|\Delta t|/\tau),
\end{equation}
where $k(\Delta t)$ measures the correlation between two observations separated by time $\Delta t$, $\sigma$ is the driving rms variability amplitude, and $\tau$ is the characteristic decorrelation (damping) timescale for returning to the mean flux. {In general, the rms variability, measured as a function of $\Delta t$ (known as the structure function; SF), can be expressed by a generalized power exponential covariance as
\begin{equation}
    {\rm SF}(\Delta t) = \SFinf \sqrt{1 - e^{-(|\Delta t|/\tau)^{\beta}}},
\label{eq:sf}
\end{equation}
where a parameter $\beta$ is introduced to describe different shapes of the SF \citep[e.g.,][]{Zu2013, Kozlowski2016_sf}.} 

In the optical regime, the SF follows a power-law form at $\Delta t \ll \tau$ with a slope of $\beta\sim1$ (i.e., a DRW process) and levels out to a constant \SFinf~at $\Delta t \gg \tau$. Its amplitude {${\rm SF}_\infty$} is found to increase with decreasing luminosity and wavelength, and the decorrelation timescale is correlated with black hole mass over $\sim6$ orders of magnitude, offering new insights into the origin of AGN variability \citep[e.g.,][]{VandenBerk2004, Morganson2014, Kozlowski2016_sf, Li2018, Burke2021, Stone2022}. 

{Compared with optical studies,} far less is known about the {variability properties} in the MIR. The MIR SF shape appears similar to that in the optical, but the power-law slope is notably steeper and the variability amplitude smaller. Moreover, its slope appears uncorrelated with AGN luminosity and wavelength, while its rms variability is anti-correlated with luminosity \citep[e.g.,][]{Kozlowski2010, Kozlowski2016, Wang2020}. These results have been qualitatively understood as the smoothing effect, in which the MIR variability is caused by reprocessing of the variable UV/optical emission, whereas the responses are diluted by the extended structure of the torus. However, a detailed study connecting the shape of the MIR SF to the torus structure and geometry is still absent from the literature.

In this paper, we investigate the ensemble variability of quasars in the optical and MIR bands, making use of optical light curves from all available ground-based surveys in the SDSS Stripe 82 region \citep[as compiled in][]{Y20} and MIR light curves from the Wide-field Infrared Survey Explorer (WISE; \citealt{Wright2010, Mainzer2011}). We will measure the ensemble SFs and directly infer the torus geometry from these second-order variability statistics. This paper is structured as follows.  The data and sample are described in Section~\ref{sec:data}. The ensemble SF measurements are presented in Section~\ref{sec:ensemble}. In Section \ref{sec:torus} we illustrate our method through simulated light curves and infer the torus parameters for SDSS Stripe 82 quasars. The conclusions are summarized in Section \ref{sec:conclusion}. Throughout this paper, timescales by default are in the restframe of the quasar. For luminosity calculations, we adopt a flat $\Lambda$CDM cosmology with $\Omega_{\Lambda}=0.7$ and $H_{0}=70\,{\rm km\,s^{-1}Mpc^{-1}}$.

\section{Sample and Data}
\label{sec:data}
To study quasar variability in the optical and MIR bands and measure the corresponding SFs, we utilize the quasar sample and their multiband light curves compiled in \cite{Y20} (Y20). The \citetalias{Y20} sample contains 587 spectroscopically confirmed SDSS quasars at $0.2\lesssim z \lesssim 2$ in the Stripe~82 region, which is selected from a parent sample of 7384 quasars by applying several cuts on variability significance and data quality\footnote{We have tested that using the entire 7384 quasars does not improve the SF measurements, given that over 80\% of the sample have MIR variability amplitudes below the noise level, making the SF measurements at short timescales noisy.}. {This sample has well measured spectroscopic properties, such as BH mass and bolometric luminosity (\lbol) from the \cite{Shen2011} catalog.} The optical light curves from various ground-based imaging surveys have an observed baseline of $\sim20$~years. The WISE light curves used in Y20 have a baseline of $\sim10$~years {(2009 Dec -- 2011 Feb, 2013 Dec -- 2019 Dec)} and a cadence of 6 months (see Table~1 in Y20 for the detailed survey information). Y20 performed cross-correlation analysis for each quasar and measured the time lag of the light curves between the $g$ band and the WISE $W1$ band. The lag is then converted to the average size of the torus {assuming} the time delay is the light-crossing time from the accretion disk to the dusty torus. Remarkably, the size of the torus follows a tight $R-L$ relation over $\sim4$ orders of magnitude in quasar luminosity when combined with local measurements of low-luminosity AGNs \citep{Kishimoto2007, Koshida2014, Lyu2019, Minezaki2019, Y20}.

In this work, we supplement the Y20 data with the latest NEOWISE 2022 data release (2020 Dec -- 2021 Dec) which extends the WISE baseline to $\sim12$~years ($\sim19$ observing epochs). Each WISE data point is the coadded {(median)} photometry of $\sim8$ observations obtained within a short $\sim1$-day window over every 6 months. {The uncertainty on the coadd photometry ($\sigma_f$) is derived by randomly perturbing each individual detection based on its photometric error (assuming a Gaussian-distributed error) within the 1-day window for 1000 times which produces 1000 median values, then use the semi-amplitude of the enclosed 16th and 84th percentiles on the sampled median to represent the $1\sigma$ uncertainty.} We measure the optical SF in the bluest $g$ band\footnote{{Y20 compiled light curves in the $g$, $r$, $i$, $V$ and unfiltered bands from various surveys and converted them into $g$ band to increase the cadence. Here we only use the actual $g$ band data.}} since it is dominated by the accretion disk continuum. {The MIR SF analysis is performed in the most sensitive $W1$ band ($3.4\, \um$) that traces the hottest dust associated with the inner edge of the torus, as well as the $W2$ band with a slightly longer wavelength ($4.6\, \um$). We will mainly explicate our methods using the $W1$ data, since its light curve is of much higher quality than that in the $W2$ band. While we also present the $W2$ results in the main text, the details of measurements in the $W2$ band are presented in Appendix \ref{appendix}. } 
We will directly compare the torus geometry probed by the SFs to that measured from dust reverberation mapping to test the feasibility of our method.

\begin{figure*}
\includegraphics[width=\linewidth]{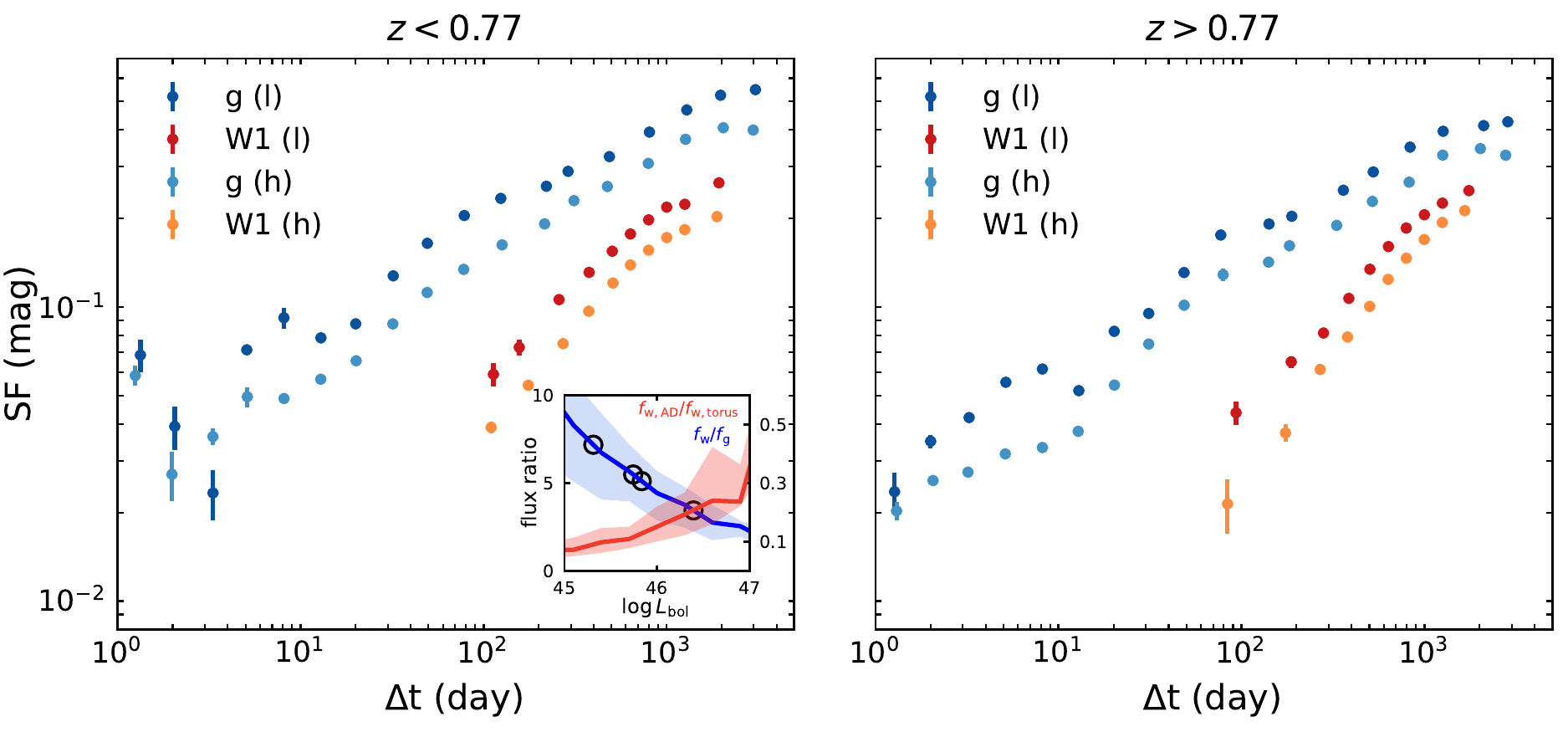}
\includegraphics[width=\linewidth]{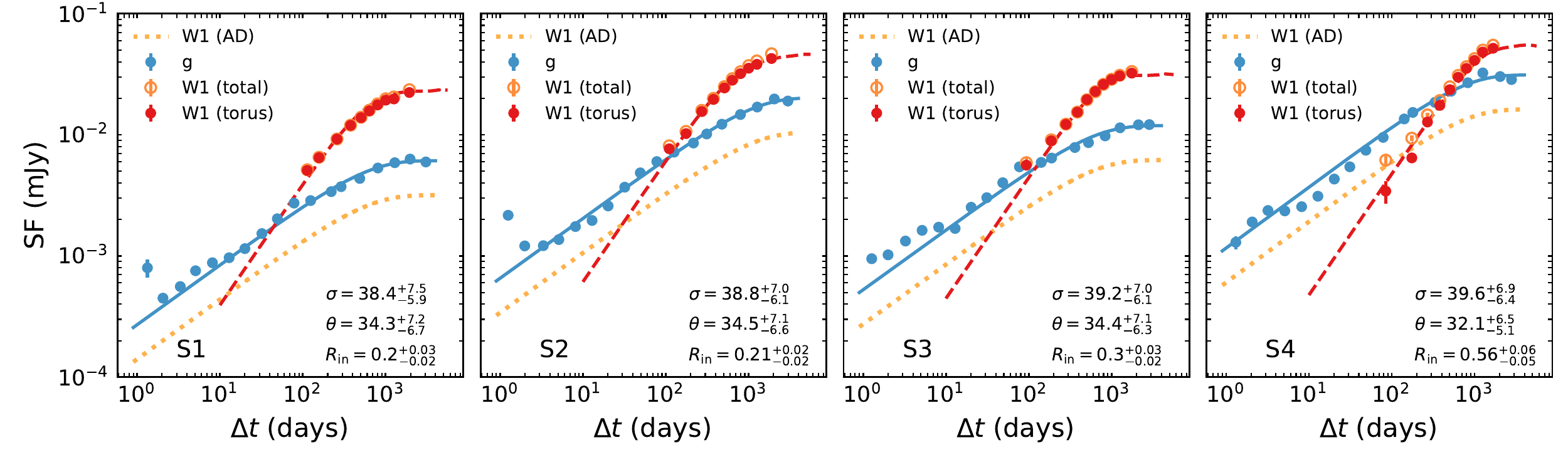}
\caption{Ensemble SFs in the $W1$ band measured in magnitude (top) and flux density (bottom) units of the 587 quasars, separated into two redshift intervals with each subsample further divided into a low luminosity (l) and a high luminosity (h) bin by their median \lbol~{(i.e., S1--S4 sample)}. In the inset of the top left panel, we also display the 16-50-84th percentiles of $f_{w}/f_g$ (left axis) and $f_{w,\rm AD}/f_{w,\rm torus}$ (right axis) as a function of luminosity. The black circles represent the median luminosity of the four subsamples.
In the bottom panels, the blue solid curves show the functional fitting results to the $g$ band SFs (blue points) using Equation~\ref{eq:sf}. The orange dotted curves show the estimated AD SFs in the $W1$ band. The orange and red points show the total (AD+torus) and the AD-subtracted (i.e., pure torus) SFs in the $W1$ band, respectively. The fitting results with our geometric torus model to the pure torus SFs are plotted as red dashed curves and the best-fit parameters are labeled. 
}
\label{fig:sf_mag}
\end{figure*}

\section{Ensemble Structure Function} 
\label{sec:ensemble}

The structure function measures the rms variability of quasars at different timescales, which is a powerful model-independent tool to empirically quantify AGN variability. Following earlier works \citep[e.g.,][]{Kozlowski2016_sf}, we measure the SF as 
\begin{equation}
{\rm SF}(\Delta t) = \sqrt{\frac{1}{N_{\Delta \rm t, pairs}} \sum_{i=1}^{N_{\rm \Delta t, pairs}}\Delta f^2},
\end{equation}
where $\Delta f$ is the magnitude or flux difference between two epochs separated by $\Delta t$ (in the rest-frame) after subtracting the photometric noise $\sigma_f$ as
\begin{equation}
\Delta f^2  = (f(t) - f(t+\Delta t))^2 - \sigma_f (t)^2 - \sigma_f (t + \Delta t)^2.
\label{eq:error}
\end{equation}
The error bars on $\rm SF(\Delta t)$ are calculated as the standard error on the mean {(i.e., std/$\sqrt{N-1}$)} for all the $\Delta f$ pairs within a given $\Delta t$ bin. {In addition, we define the observed noise term (\nerr) as the mean of all $\sigma_f (t)^2 + \sigma_f (t + \Delta t)^2$ then take the square root.} 


Theoretically, the SF can be measured for light curves at arbitrary observing segments assuming that AGN variability is a stationary process, and should be insusceptible to sampling problems. However, in reality, the AGN variability may be non-stationary, or even if it is stationary, the baseline of the survey is not long enough to well constrain the flattening part of the SF at long timescales \citep[e.g.,][]{Kozlowski2017}. The sparse and/or irregular time sampling will also cause spurious features in the SFs owning to the lack of sufficient pairs to robustly measure an average $\Delta f$ \citep[e.g.,][]{Emmanoulopoulos2010}. These problems can be partially alleviated by measuring the {\em ensemble SF} which is the average of the SFs of many quasars with similar properties \citep[e.g.,][]{VandenBerk2004}. 

{It is important to emphasize that the accuracy of noise subtraction can significantly affect the ensemble SF measurement. While Equation~\ref{eq:error} provides a straightforward approach for subtracting the noise (hereafter the \nerr~method), its accuracy strongly depends on the accuracy of the reported uncertainties of photometric measurements. Inaccurate estimation of magnitude/flux uncertainties could introduce biases in the shape of the SF. In fact, our analysis reveals that the \nerr~method may result in systematic overestimation of the noise term for the $W2$ band, which has lower light curve quality relative to the $W1$ band (see Appendix \ref{appendix} for details).
Alternatively, a functional fitting method can be used to estimate the noise term. As we will demonstrate in Section~\ref{subsec:simu}, Equation~\ref{eq:sf} can adequately describe the MIR SF at timescales of $100 \lesssim \Delta t \lesssim 2000$~days. By adding a constant noise term (\cerr) in quadrature to Equation \ref{eq:sf} and fitting the observed ensemble SF (without subtracting the photometric noise beforehand), which flattens at short timescales due to flux errors (i.e., the SF floor; \citealt{Kozlowski2016_sf}), the noise term and its associated uncertainty can be determined (hereafter the \cerr~method).} 

{We note that the \cerr~method is applicable only if the noise term is nearly constant with time, which is the case for the WISE light curves. In contrast, the $g$ band light curves are obtained from multiple surveys conducted during various observing seasons. They contribute differently to the SF at different timescales, leading to a notable time variation in the noise term. Therefore, we always use Equation~\ref{eq:error} to subtract the noise in the $g$ band\footnote{{Our results are insensitive to the shape of the $g$ band SF at $\Delta t \lesssim 10$~days where noise subtraction impacts the SF the most. }}.
The noise terms estimated from the \nerr~and \cerr~methods show excellent agreement in the $W1$ band. For the sake of simplicity, we present our methods and results using the $W1$ SFs measured from the \nerr~method, which is straightforward to compute and consistent with our treatment for the $g$ band. However, we report the $W2$ results based on the \cerr~method as default since the \nerr~method likely overestimates the photometric noise. By subtracting the constant SF floor with the \cerr~method, we remove all contributions from flux uncertainties in the SF measurements. We refer the readers to Appendix \ref{appendix} for a detailed comparison of the results obtained from different noise estimation methods and the discussion of the impact of imperfect noise subtraction.}

To measure the ensemble SF, we split our quasar sample into two redshift bins, $z < 0.77$ and $z>0.77$, based on the median redshift, and further divide each subsample by their median bolometric luminosity (\lbol). This sample division provides a higher-$L$ subsample and a lower-$L$ subsample in a given redshift bin, where the WISE bands trace the same rest-frame infrared wavelength. As a result, we have four subsamples each containing $\sim150$ quasars. The median redshift and luminosity for each samples are $z \approx 0.50$, 0.64,  0.93, 1.20 and $\loglbol \approx 45.3$, 45.7, 45.9 and 46.4~\ergs, respectively. These four subsamples will be referred to as S1, S2, S3, and S4, respectively.

In the top panels of Figure \ref{fig:sf_mag}, we present the ensemble SFs {in the $W1$ band} measured in magnitude units, as commonly adopted in the literature, to qualitatively show their differences in the optical and MIR bands. The uncertainties on the ensemble SFs are calculated by bootstrapping each individual SF and deriving the semiamplitude of the enclosed 16th and 84th percentiles on the bootstrapped means. The $W1$ SFs increase with $\Delta t$ in a way that resembles the shape of the optical SFs. Additionally, the variability amplitudes decrease with increasing \lbol~at fixed redshift in both bands, consistent with the scenario that the MIR variability is caused by reprocessing of UV/optical emission. However, it can be clearly seen that the $W1$ SFs are steeper than that in the $g$ band and have smaller amplitudes at all timescales being probed. 

A functional fit to the $W1$ SFs using Equation~\ref{eq:sf} yields a short-timescale slope of $\beta\approx1.3-2.0$ (Table \ref{table:fit}), which is much steeper than that reported in \cite{Kozlowski2016} based on Spitzer light curves at $3.6\,\um$ and $4.5\,\um$ of $1<z<3$ AGNs ($\gamma \approx 0.45$ where $\beta\sim2\gamma$). Moreover, the slope appears to increase with \lbol, while no correlation was found in \cite{Kozlowski2016}. This could be due to the insufficient data quality in \cite{Kozlowski2016} since their Spitzer light curves only cover five epochs with large gaps (several years) exist between adjacent observing windows. {Analogous findings are observed in the $W2$ band, with the slope and variability amplitude being slightly steeper and smaller, respectively, when compared to the $W1$ band (Appendix \ref{appendix})}.

\begin{table}
\renewcommand{\arraystretch}{1.2}
\caption{Model fitting results in the form of Equation~\ref{eq:sf} to the measured SFs.}
\centering

$g$ band SF measured in magnitude unit
\begin{tabular}{ccccc}
\hline
\hline
redshift & \loglbol & \SFinf~(mag) & $\tau\ \rm (days)$ & $\beta\ \rm (fixed)$ \\
\hline
$z<0.77$ & 45.3 & $0.544_{-0.002}^{+0.002}$ & $853.1_{-11.3}^{+12.0}$ & 1.0\\
$z<0.77$ & 45.7 & $0.414_{-0.001}^{+0.001}$ & $856.7_{-6.7}^{+7.0}$  & 1.0\\
$z>0.77$ & 45.9 & $0.427_{-0.001}^{+0.001}$ & $725.4_{-4.9}^{+4.8}$ & 1.0\\
$z>0.77$ & 46.4 & $0.366_{-0.001}^{+0.001}$ & $952.1_{-7.1}^{+7.3}$ & 1.0\\
\hline\\
\end{tabular}

Total SF in the $W1$ band measured in magnitude unit
\begin{tabular}{ccccc}
\hline
\hline
redshift & $\loglbol$ & $\SFinf$~(mag) & $\tau~\rm (days)$ & $\beta$ \\
\hline
$z<0.77$ & 45.3 & $0.278_{-0.007}^{+0.008}$ & $1080.2_{-78.8}^{+96.8}$ & $1.31_{-0.05}^{+0.05}$\\
$z<0.77$ & 45.7 &
$0.206_{-0.002}^{+0.002}$ & $895.0_{-29.0}^{+33.0}$ & $1.59_{-0.04}^{+0.04}$\\
$z>0.77$ & 45.9 & $0.259_{-0.005}^{+0.006}$ & $995.3_{-48.9}^{+59.3}$ & $1.69_{-0.06}^{+0.05}$\\
$z>0.77$ & 46.4 & $0.226_{-0.005}^{+0.006}$ & $1097.5_{-49.5}^{+56.9}$ & $1.90_{-0.06}^{+0.06}$\\
\hline\\
\end{tabular}

$g$ band SF measured in flux unit
\begin{tabular}{ccccc}
\hline
\hline
redshift & \loglbol & \SFinf~(mJy) & $\tau\ \rm (days)$ & $\beta\ \rm (fixed)$ \\
\hline
$z<0.77$ & 45.3 & $0.006_{-0.000}^{+0.000}$ & $530.9_{-5.1}^{+5.1}$ & 1.0\\
$z<0.77$ & 45.7 & $0.020_{-0.000}^{+0.000}$ & $998.0_{-9.9}^{+10.6}$ & 1.0\\
$z>0.77$ & 45.9 & $0.012_{-0.000}^{+0.000}$ & $537.4_{-3.7}^{+3.8}$ & 1.0\\
$z>0.77$ & 46.4 & $0.034_{-0.000}^{+0.000}$ & $897.6_{-6.5}^{+6.6}$ & 1.0\\
\hline\\
\end{tabular}

Pure-torus SF in the $W1$ band measured in flux unit
\begin{tabular}{ccccc}
\hline
\hline
redshift & $\loglbol$ & $\SFinf\ \rm (mJy)$ & $\tau~\rm (days)$ & $\beta$ \\
\hline
$z<0.77$ & 45.3 & $0.023_{-0.000}^{+0.000}$ & $850.5_{-38.0}^{+43.7}$ & $1.41_{-0.04}^{+0.04}$\\
$z<0.77$ & 45.7 & $0.044_{-0.001}^{+0.001}$ & $941.6_{-41.7}^{+47.6}$ & $1.62_{-0.04}^{+0.04}$\\
$z>0.77$ & 45.9 & $0.033_{-0.000}^{+0.001}$ & $833.7_{-30.0}^{+34.0}$ & $1.67_{-0.05}^{+0.05}$\\
$z>0.77$ & 46.4 & $0.053_{-0.001}^{+0.001}$ & $987.2_{-26.4}^{+30.0}$ & $2.20_{-0.05}^{+0.06}$\\
\hline\\
\end{tabular}

\label{table:fit}
\end{table}

The accretion disk (AD) could also contribute to the observed emission at rest-frame $\sim1-2\ \um$ and contaminate the torus variability measurements \citep[e.g.,][]{Kishimoto2007, Kishimoto2008, Lira2011}. This is illustrated in the inset of Figure \ref{fig:sf_mag} where we show that the $W1$ to $g$ band flux ratio ($f_w/f_g$) of our sample (blue line) decreases significantly with \lbol. Assuming the SED of the AD follows the $F_\nu \propto \nu^{1/3}$ power law that extends to $\lambda_{\rm rest} \sim2\,\um$ \citep{Kishimoto2008}, we estimate the AD flux in the $W1$ band ($f_{w,\rm AD}$) from the $g$ band flux and show its relative strength to the pure torus flux ($f_{w,\rm torus}$) in the inset of Figure \ref{fig:sf_mag} (red line). 
The non-negligible value\footnote{The contamination from the host galaxy has not been taken into account. The actual $f_{w,\rm AD}/f_{w,\rm torus}$ for the brightest quasars at $z>1$ could be even higher as the host galaxies have a stronger contribution at $\lambdar \sim 1.5\,\um$ compared to $\lambdar \sim 2000\,\rm \AA$.} at the highest luminosities suggests that the AD could contribute significantly to the short-timescale variability in the $W1$ band. {Conversely, given the rising shape of the torus SED and the declining trend of the AD SED in the MIR, the AD contamination in the $W2$ band is insignificant.}

Considering the sparse time sampling of the light curves and the absence of $g$ band observations after 2018, we adopt a statistical approach to remove the AD contamination, instead of subtracting the AD flux from each IR light curves as done in earlier works \citep[e.g.,][]{Koshida2014, Lyu2019}. The idea is to measure the SF in flux density unit (weighted by luminosity distance$^2$ for each object in the bin to account for the redshift effect; see bottom panels in Figure \ref{fig:sf_mag}) and infer the SF of the disk continuum in the WISE bands (\SFwdisk) from that in the $g$ band (\SFg), then subtract it from \SFw~to obtain the pure torus component (\SFwtorus). Other sources of contamination to the WISE flux, such as a constant host galaxy or nearby objects will also be subtracted out when measuring $\Delta f$ in flux. 

The details and feasibility of our subtraction methodology will be further illustrated in Section \ref{subsec:disk} with simulated light curves and SFs. Here we briefly outline our method. 
Given the weak wavelength dependence of AD variability \citep[e.g.,][]{Stone2022}, \SFwdisk~can be estimated as $\SFg \times (\nu_w/\nu_g)^{1/3}$. If we roughly assume that the flux variations at different timescales follow a Gaussian distribution and associate its standard deviation with the rms variability, then \SFwtorus~can be derived by solving 
\begin{equation}
    \SFw^2 = \SFwtorus^2 + \SFwdisk^2 + 2\rho \times \SFwtorus\, \SFwdisk,
\label{eq:torus_sf}
\end{equation}
where $\rho$ is the covariance (correlation coefficient) between $f_{w,\rm AD}$ and $f_{w,\rm torus}$ as a function of timescale. In practice, this assumption does not hold since AGN variability is red-noise like instead of a Gaussian white noise. However, we will demonstrate in Section \ref{subsec:disk} that it provides a reasonable approximation. The value of $\rho$ depends on the geometry of the torus: an extremely compact (extended) torus with $R\rightarrow 0$ ($R\rightarrow \infty$) will have $\rho \rightarrow 1$ ($\rho \rightarrow 0$). Observationally, it is challenging to accurately constrain $\rho$ with our sparse light curves since the optical and WISE data do not overlap well. Therefore, we adopt $\rho$ measured from simulated light curves which is typically $-0.3\sim 0$ for a broad range of torus geometry at timescales of $\sim100-2000$~days (see Section \ref{subsec:disk} for details). 

The resulting pure-torus SFs in the $W1$ band are shown in Figure \ref{fig:sf_mag} (bottom panels). This time the amplitudes of the MIR SFs are greater than those in the optical as the absolute flux in the $W1$ band is typically a factor of few times larger than that in the $g$ band. The AD contamination is small overall, except for the most luminous S4 sample, where its variability in the $W1$ band is comparable to the torus at $\Delta t \sim 100$~days.
The functional fitting results with Equation~\ref{eq:sf} to the $g$ band ($\beta$ fixed to 1.0, i.e., assuming a DRW process) and pure-torus SFs measured in flux unit are summarized in Table~\ref{table:fit}. The $g$ band SFs can be well described by a DRW model with $\tau\sim700$~days, {consistent with previous dedicated studies on the AGN optical variability \citep[e.g.,][]{Stone2022}}. The $W1$ slope becomes slightly steeper compared to that measured in magnitude units {because of the subtraction of the AD contamination}, while its correlation with luminosity remains.  

The steeper MIR SF demonstrates that the variability in the MIR is suppressed compared to the optical variability, and the suppression is most prominent at short timescales. This could be attributed to the smoothing effect, where the optical variations of the AD have been smeared out by the extended structure of the torus, due to that different parts of the torus respond asynchronously to the incident UV/optical continuum. The optical and MIR SFs thus contain critical information on the geometry of the torus (e.g., inclination, size, opening angle). For example, consider a ring-like torus with a face-on inclination. In this configuration, the observed MIR light curve would simply be a time-delayed version of the optical light curve, and the shapes of their SFs should be identical. Therefore, the torus probed by the WISE bands must have some vertical or radial extension and/or the inclination angle is non-zero. Higher-luminosity quasars may have a more extended dust distribution in order to produce a steeper $\beta$.
In the next section, we will generate simulated light curves using a toy torus model to test this hypothesis.

\section{Constraining Torus Geometry}
\label{sec:torus}

\subsection{Idealized Light Curves}
\label{subsec:simu}

\begin{figure}
\includegraphics[width=\linewidth]{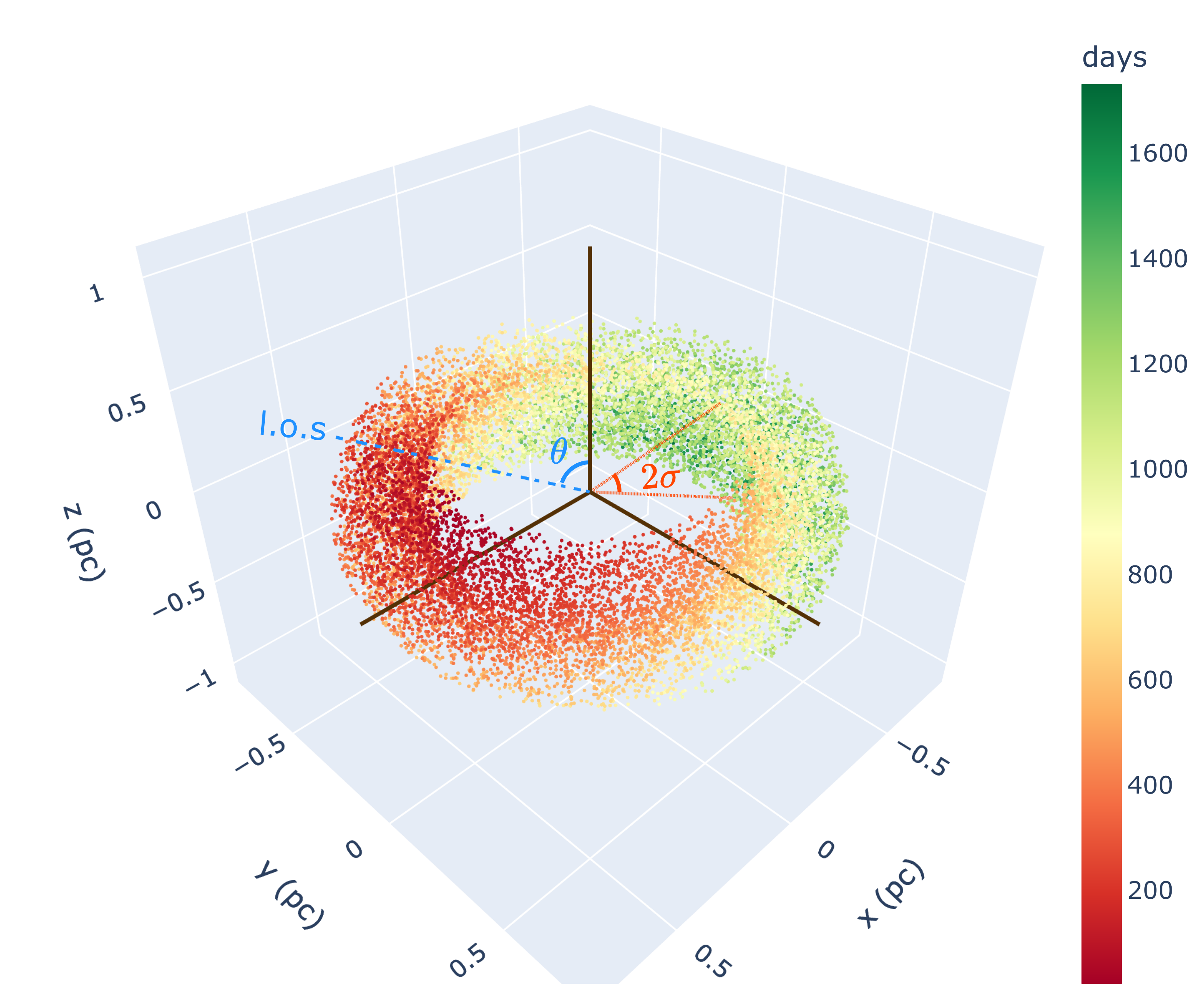}
\caption{A schematic view of the torus model with $\Rin=0.5$~pc, $\theta=45^\circ$, $\sigma=30^\circ$, $Y=1.5$, and $p=-1.0$. Each cloud is color-coded by the time delay relative to the observer's line-of-sight, indicated by the vertical color bar.}
\label{fig:torus}
\end{figure}

\begin{figure*}
\centering
\includegraphics[width=\linewidth]{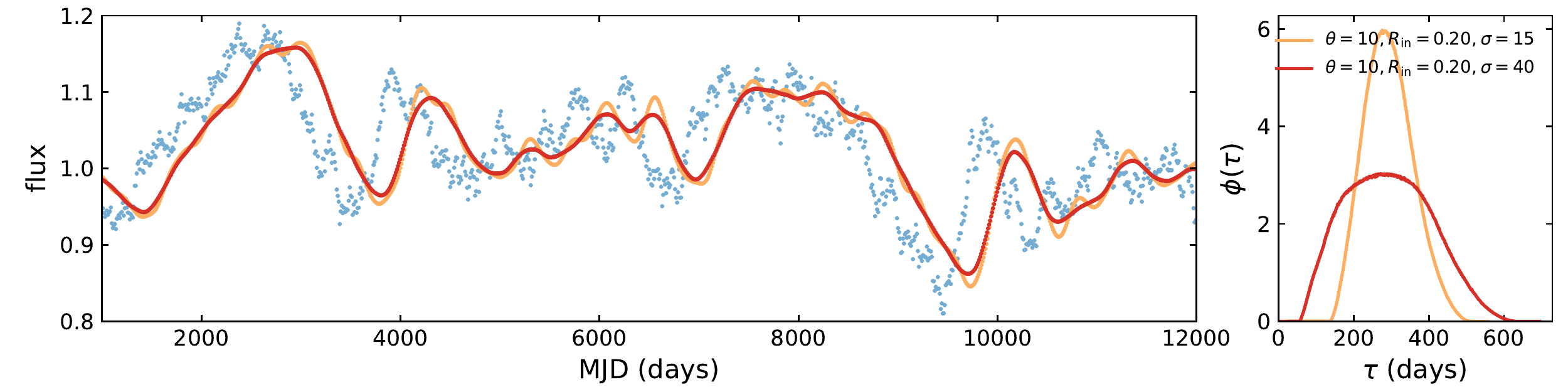}
\includegraphics[width=\linewidth]{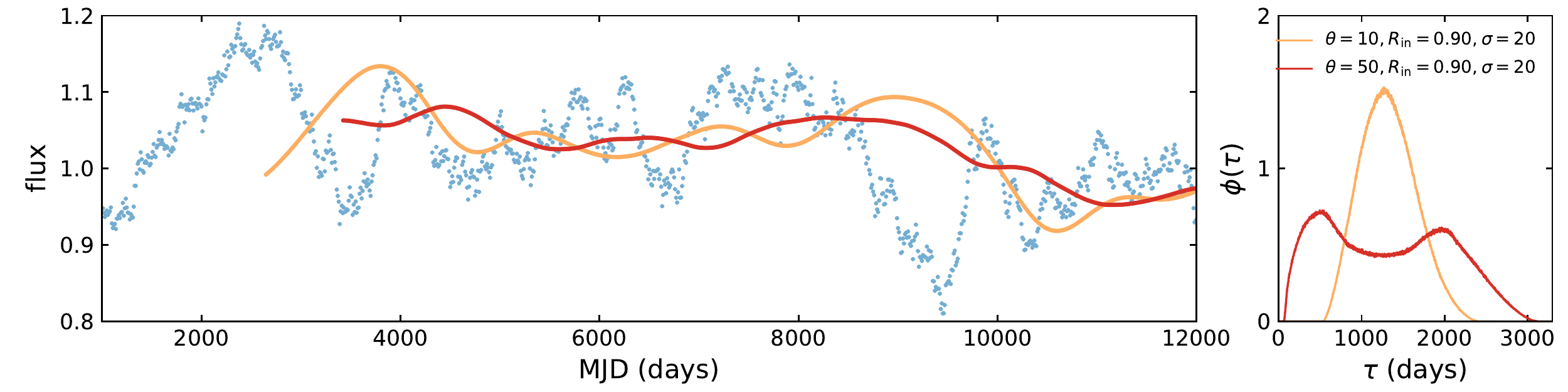}
\caption{Simulated idealized light curves (left panels) with a 5-day cadence in both $g$ band (blue) and $W1$ band (orange and red) for different torus geometries. The transfer functions and parameters for the torus models are shown in the right panels. }
\label{fig:simu_lc}
\end{figure*}

\begin{figure*}
\centering
\includegraphics[width=\linewidth]{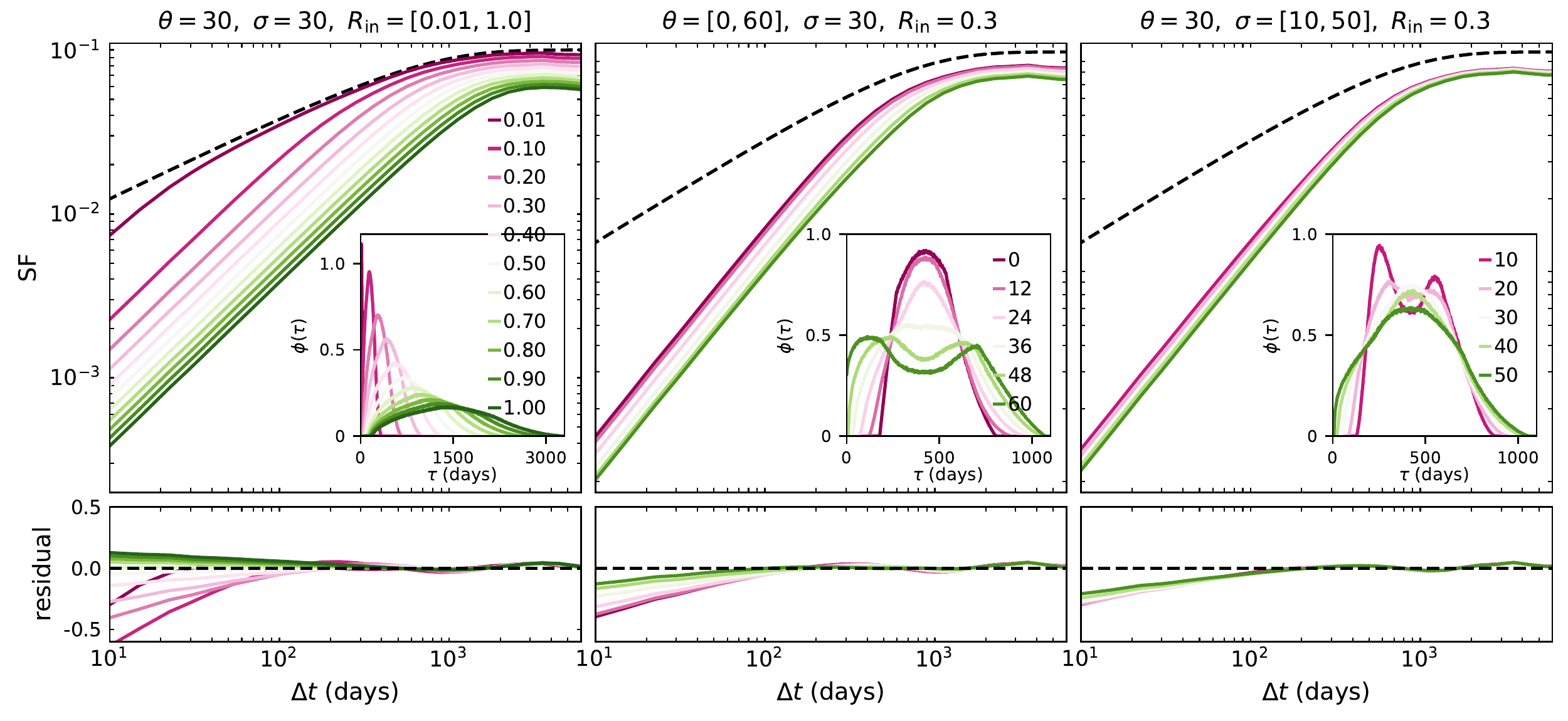}
\caption{Top: SFs of the simulated light curves in the optical (dashed) and MIR (solid) for different torus parameters as indicated in the figure. The curves from top to bottom correspond to increasing \Rin, $\theta$ and $\sigma$. The transfer functions of each torus models are shown in the inset (arbitrary normalization). Bottom: {normalized residual of fitting each SF in the top panels with Equation \ref{eq:sf} (i.e., fitting residual/SF).} }
\label{fig:simu_sf}
\end{figure*}

We first generate idealized light curves (regularly sampled with a 5-day cadence and a baseline of $\sim10^4$ days {in both bands}) to illustrate the shape of the MIR SF under different torus geometries. The simulated light curves and the corresponding SFs in the following sections are presented in flux units (arbitrary normalization). The model parameters are summarized in Table~\ref{table:params}. Below we describe the models and parameters in detail.

\begin{table}
\renewcommand{\arraystretch}{1.1}
\caption{Parameters used to generate simulated optical and MIR light cuves.}
\centering
\begin{tabular}{cc}
\hline
\hline
parameter & description\\
\hline
$\SFinf$ & optical $\SFinf$ in Eq.~\ref{eq:sf}\\
$\tau$ & optical damping timescale (day) in Eq.~\ref{eq:sf}\\
$\beta$ & optical slope in Eq.~\ref{eq:sf}; fixed to 1.0\\ 
$f_{g, \rm AD}$ &  mean flux in the optical; fixed to 1.0\\
$\theta$ & inclination angle; $\theta=0^\circ$ means face-on\\
$\Rin$ & torus inner radius (pc) \\
$Y$ & ratio of outer to inner radius; fixed to 1.5\\
$\sigma$ & half-opening angle (degree)\\
$p$ & radial powerlaw index; fixed to --1.0\\
$f_{w,\rm torus}/f_{g, \rm AD}$ & mean MIR/optical flux ratio\\
\hline\\
\end{tabular}
\label{table:params}
\end{table}

The optical light curve is generated from a DRW model using {\tt{celerite}} \citep{celerite} {assuming a mean flux of 1.0, a $\sim10\%$ fractional variability amplitude ($\SFinf=0.1$)} and a decorrelation timescale $\tau=700$ days, as motivated by observations (Table \ref{table:fit}).  The MIR light curve is the convolution of the optical light curve with the torus transfer function \trans~as
\begin{equation}
F_{\rm MIR} (t) = \famp \int F_{\rm optical} (t-\tau)\, \trans d\tau,
\end{equation}
where \trans~quantifies the delayed response (at time lag $\tau$) of distant dusty clouds to the continuum emissions from the accretion disk, and \famp~is the ratio between the variable part of the optical and IR emissions which is related with the dust amount and  reprocessing efficiency. The transfer function depends on the geometry, composition (grain species), and physics (e.g., anisotropic illumination, optical thickness, cloud occultation) of the torus \citep[e.g.,][]{Kawaguchi2011, Almeyda2020}, while a simplified top-hat transfer function is often employed in dust lag measurements \citep[e.g.,][]{Lyu2019, Y20}.

In this work, we consider a simple geometric model and neglect detailed radiative transfer given that the light curve quality is insufficient to constrain the complicated physics that could be highly degenerate. Our torus model contains five parameters: the inclination angle ($\theta = 0^\circ$ means a face-on view), the inner radius \Rin\ (in units of pc), the outer to inner radius ratio $Y$, the torus half-opening angle $\sigma$, and the radial power-law index $p$. Specifically, since we are measuring the ensemble SF in a narrow wavelength range {($\lambdar \sim1.4-2.8\,\um$ in $W1$ and $\lambdar \sim2.0-3.8\,\um$ in $W2$)}, the clouds are expected to be the hottest dust concentrate near the sublimation radius ($T\sim1500$~K, $R\lesssim 1$~pc), as opposed to the polar dust component ($T\sim100-200$~K, $R\sim100-1000$~pc) that dominates the emissions at $\lambda \sim20\,\um$ and contribute little to the $W1$ and $W2$ bands \citep[e.g.,][]{Honig2013, Lyu2018, Yamada2023}. Therefore, we assume that the torus has a thin-wall geometry by fixing $Y$ to a small value of 1.5\footnote{Assuming $T_{\rm in}=1500$~K (i.e., the sublimation temperature for Silicate) and a $T\propto R^{-1}$ radial temperature profile \citep{Kishimoto2011}, $Y\sim T_{\rm in}/T_{2.8\,\um} \sim$~1500~K/1000~K according to  Wien's displacement law.}, and neglect the polar component in our model.  In the radial direction, the clouds are distributed following a power-law profile with its index being fixed to --1.0\footnote{We find that as long as $Y$ is small (e.g., $\lesssim2.0$), the choice of $p$ has a negligible impact on the shape of the transfer function.} \citep[e.g.,][]{Kishimoto2011, Lyu2019}. In the vertical direction, the clouds have a sharp edge at [$-\sigma$, $\sigma$] and within $\sigma$ they follow a Gaussian distribution with the standard deviation being set to $\sigma$. As a result, most clouds are clustered on the equatorial plane.
In addition, we define the characteristic (response-weighted) time delay, comparable to the time lag measured from dust reverberation mapping, as
\begin{equation}
    \tau_{\rm lag} = \frac{\int{\tau \phi(\tau) d\tau}}{\int{\phi(\tau)d\tau}},
\end{equation}
and convert it to the effective size of the torus (\Reff) by multiplying the speed of light. A schematic view of our torus model is shown in Figure \ref{fig:torus}. 

We trace the time delay of each cloud (Figure \ref{fig:torus}) relative to the observer's line of sight and measure the transfer function for different torus geometries. In Figure \ref{fig:simu_lc} we show four examples of the transfer functions and the resulting MIR light curves assuming the average MIR to optical flux ratio ($f_{w,\rm torus}/f_{g, \rm AD}$) to be 1.0. For a compact torus with $\Rin=0.2$~pc, the MIR light curve follows the general variability pattern in the optical with a clear time delay; while for an extended torus with $\Rin=0.9$~pc, the variability in the MIR is significantly diluted. Increasing $\sigma$ and $\theta$ can further smear out the variability features. 

In Figure \ref{fig:simu_sf} we plot the SFs of the simulated light curves which are derived from 100 ensembles to mitigate random fluctuations. In the first panel we demonstrate how the SF changes with \Rin~under fixed $\theta$ ($30^\circ$) and $\sigma$ ($30^\circ$). At $\Rin=0.01$~pc, the MIR SF follows that in the optical at long timescales. However, its slope becomes steeper and resembles a power-law shape below a timescale of $\sim100$~days, meaning that the rapid flux variations are suppressed. As \Rin~increases, the dilution effect further weakens the short-timescale variability and starts to impact the longer-timescale regime. 
When focusing on $100 \lesssim \Delta t  \lesssim 2000$~days (i.e., the timescale probed by observations), the slope $\beta$ becomes steeper as \Rin~increases. This trend resembles the correlation between $\beta$~and \lbol~we found in Section \ref{sec:ensemble}, indicating that the size of the torus may increase with \lbol. In the second and third panels we fix \Rin~and vary $\theta$ and $\sigma$. Increasing $\theta$ and $\sigma$ has a similar effect on the shape of the SF as increasing \Rin, meaning that there are inherent degeneracies among different torus parameters.

\begin{figure}
\includegraphics[width=\linewidth]{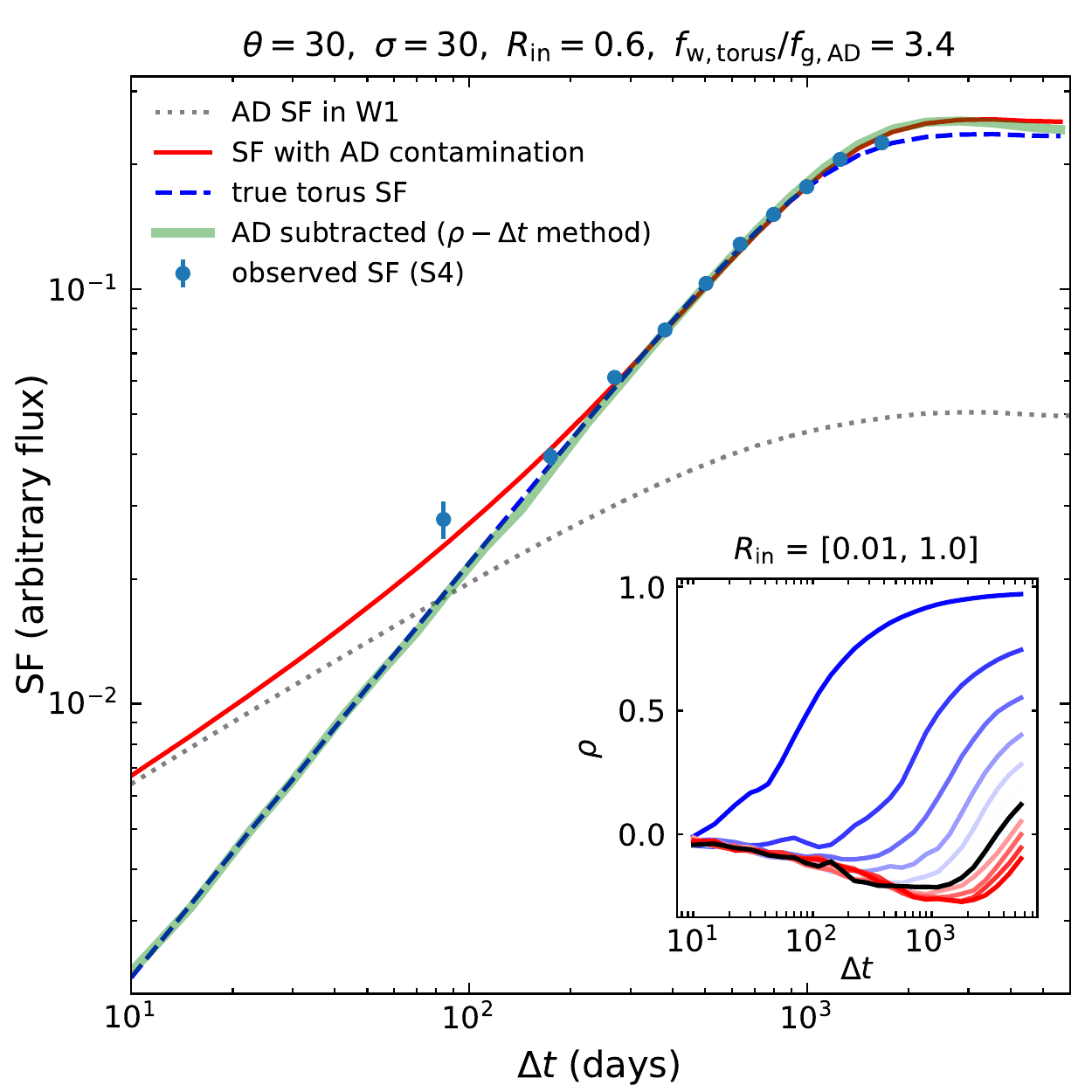}
\caption{The impact of accretion disk contamination on the SF measurement in the $W1$ band for a simulated disk-torus system with $\theta=30$, $\sigma=30$, $\Rin=0.6$, and $f_{w,\rm torus}/f_{g, \rm AD}=3.4$.  The blue dashed, gray dotted, and red solid curves represent the SF of the torus, disk, and torus+disk in the $W1$ band, respectively. The disk-subtracted $W1$ SF is shown in green. The $W1$ SF of the S4 sample is shown as blue points.
The covariance between optical and MIR light curves as a function of timescale is shown in the inset for $\Rin=[0.01,\ 1.0]$. The curves from top to bottom correspond to increasing $\Rin$ with a step of 0.1; $\Rin=0.6$ is highlighted in black.}
\label{fig:AD}
\end{figure}

It is worth noting that the MIR SFs do not strictly follow Equation \ref{eq:sf}. {This empirical functional form fails to decribe the shape of the model SF at the shortest timescales, especially for compact torii where the suppression of short timescale variability is stronger than that predicted by Equation \ref{eq:sf}.} This is illustrated in the bottom panels of Figure \ref{fig:simu_sf} where we show that the normalized residuals of fitting the model SFs with Equation \ref{eq:sf} can be as large as $\sim50\%$ at $\Delta t \lesssim 100$~days. However, it can still provide a good fit to the data at $\Delta t \gtrsim 100$~days ({i.e., the timescales probed by our WISE light curves}), thus offering a convenient means to quantify the observed long-timescale variability. 

Overall, the shape of the simulated SFs are in good agreement with those of SDSS quasars, lending support that the observed MIR variability is originated from the delayed and smoothed response to the UV/optical variations. Furthermore, the MIR SFs of SDSS quasars at $\Delta t \sim 100-600$ days follow a steep power-law shape, distinct from the model SFs with $\Rin<0.1$~pc. This result disfavors a compact torus geometry for these quasars.

\subsection{Effect of Accretion Disk Contamination}
\label{subsec:disk}
As noted in Section \ref{sec:ensemble}, the variability of disk continuum in the MIR can affect the SF measurement for the most luminous quasars. To demonstrate this effect, we generate a simulated $W1$ light curve with $\Rin=0.6$, $\theta=30^\circ$, $\sigma=30^\circ$, $f_{w,\rm torus}/f_{g, \rm AD}=3.4$, and add disk contamination represented by $f_{w,\rm AD} = f_{g,\rm AD}\times(\nu_w/\nu_g)^{1/3}$. The parameters are selected to be representative of luminous quasars ($\loglbol \sim 46.4)$ given the $f_{w,\rm torus}/f_{g, \rm AD}-\lbol$ relation shown in Figure \ref{fig:sf_mag} and the $R\propto L^{0.5}$ relation found in the literature \citep[e.g.,][]{Koshida2014, Lyu2019}.
The resulting SFs with and without AD contamination are displayed in Figure \ref{fig:AD}. A noticeable leveling off towards $\Delta t \lesssim 300$~days is evident when the disk contamination is included in the MIR SF measurement; there is also an excess relative to the true SF at $\Delta t \gtrsim 1000$~days. The flattening at short timescales can also be seen in the observed SF, demonstrating that its shape is a superposition of the torus and AD component. Nevertheless, the variability signals at the timescales probed by our light curves are dominated by the torus emission.

To remove the disk contamination using the method outlined in Section \ref{sec:ensemble}, we first measure $\rho$ as a function of $\Delta t$ from our idealized light curves (shown as a black curve in the inset of Figure \ref{fig:AD}). The pure-torus component, \SFwtorus, is then calculated using Equation \ref{eq:torus_sf}. The AD-subtracted SF is in good agreement with the true model (Figure \ref{fig:AD}), demonstrating the effectiveness of this method.  The slight overestimate ($\sim5\%$) of the true SF at the longest timescales may be attributed to the significant deviation of AGN variability from a Gaussian process as assumed in Section~\ref{sec:ensemble}. 

When applying this method to observed quasars, one can roughly estimate \Rin~from the $R\propto L^{0.5}$ relation and derive a dedicated $\rho-\Delta t$ relation. Figure \ref{fig:AD} shows examples with $\Rin = [0.01\sim1.0, dR=0.1]$ in the inset. It can be seen that the $\rho-\Delta t$ relation is insensitive to $\Rin$ (as well as $\theta$ and $\sigma$, which are not shown) at $\Delta t \sim 100-1000$ days for $\Rin \gtrsim 0.4$. This critical size corresponds to $\loglbol \gtrsim 46.0$, consistent with our high luminosity samples. On the other hand, a compact torus with $\Rin \lesssim 0.4$ is expected to be associated with low luminosity AGNs, and thus, subtracting the AD contamination or not will have a minimal impact on the SF.

\subsection{Realistic Light Curves}
\label{subsec:params}

The idealized simulations presented in Section \ref{subsec:simu} have demonstrated that the geometry of the torus is imprinted in the shape of the MIR SF, and its difference compared with the optical SF. In principle, this could be used to constrain torus parameters by fitting the measured ensemble SFs to models. However, in practice, the sparse time sampling of the light curve, photometric noise, and limited number of ensembles make it impossible to measure the SF with the precision like the models shown in Figure \ref{fig:simu_sf}. Additionally, the observed ensemble SF is the average of individual SFs associated with a broad range of torus parameters. It is unclear whether the average SF also corresponds to the average of the underlying torus parameter distributions.
Here we simulate realistic light curves and test to what extend can the current data be used to constrain the geometry of the torus. 

We first summarize the observational constraints on the distribution of torus parameters that will be used to generate our simulated populations. The effective size of the torus at $\lambdar \sim 2\,\um$ is $\sim0.1-1.0$ pc for luminous quasars. At a given luminosity, it follows a log-normal distribution with an intrinsic scatter of only $\sim0.1$ dex \citep{Koshida2014, Y20}. The opening angle $\sigma$ can be derived from $f_{\rm IR}\equiv L_{\rm IR}/\lbol$ once corrected for the anisotropic effect as detailed in \cite{Stalevski2016}, and is typically $30^\circ - 50^\circ$ for Type 1 quasars \citep[e.g.,][]{Ezhikode2017, Ichikawa2019}. The inclination angle should be randomly distributed within [$0^\circ$, $90^\circ-\sigma$] according to the unification model, and its average value is expected to be $\sim25^\circ-45^\circ$. This simple argument is consistent with inclinations constraint from SED fitting, kinematic modeling and polarization observations \citep[e.g.,][]{Fischer2013, Marin2014, Zhuang2018}.

Based on these results, {we consider a hypothetical population of Type 1 quasars that varies as a DRW process in the optical band.} The decorrelation timescale $\tau$ (in rest-frame days) is assumed to follow a Gaussian distribution $\mathcal{N}(600\sim700,\ 50)$, and $\SFinf \sim \mathcal{N}(0.20\sim0.25,\ 0.02)$. For a mean optical flux of 1.0 (which will be used in our simulations), this \SFinf~corresponds to a fractional variability amplitude of $\sim16\%$, similar to the observed value for our sample.
The torus opening angle, inclination angle, and logarithm of $\Rin$ are modeled as Gaussian-distributed with $\sigma \sim \mathcal{N}(30\sim50,\ 10)$, $\theta \sim \mathcal{N}(25\sim45,\ 10)$, and ${\rm log}\Rin \sim \mathcal{N} (-1.0 \sim -0.1,\ 0.1)$, respectively. 

In our simulation, we first generate $\overline{\tau}$, $\overline{\SFinf}$, $\overline{\sigma}$, $\overline{\theta}$, and $\overline{{\rm log}\,\Rin}$ from a uniform distribution $U(600, 700)$, $U(0.20, 0.25)$, $U(30, 50)$, $U(25, 45)$, and $U(-1.0, -0.1)$, respectively, which we treat as the average properties of the hypothetical quasar populations. Then for each set of average parameters, we create a simulated quasar sample and assign each quasar an optical variability pattern and torus geometry by randomly sampling the aforementioned Gaussian distributions for 150 times. The sample size is chosen to match our S1--S4 samples.
Next, for each simulated quasar, we generate light curves with realistic photometric uncertainties (0.02 for $g$ band and 0.04 for $W1$ band) and cadence (by using the observing epochs of randomly selected SDSS quasars). An ``observed'' ensemble SF (\SFobs) {similar to those shown in Figure \ref{fig:sf_mag}} is then derived for the 150 simulated light curves in both bands.

{The next step is to generate a model SF  library in the MIR. We first fit the $g$ band \SFobs~with Equation~\ref{eq:sf} and obtain the decorrelation timescale $\tau_{\rm fit}$, which is typically $500-800$ days. The MIR model SFs used in the fitting are then generated from a grid of parameters spanning $\tau=$ the nearest integral multiple of 100 to $\tau_{\rm fit}$ (in the optical), $\SFinf=0.1$ (optical), $\theta = [25\sim45,\ d\theta=1]$, $\Rin = [0.01\sim1.0,\ dR=0.01]$, and $\sigma=[30\sim50,\ d\sigma=2]$ following the method described in Section \ref{subsec:simu}. A parameter $f_{\rm norm}$ is introduced in the fitting to account for the relative normalizations.}
By searching for model SF {(i.e., example curves displayed in the top panels of Figure \ref{fig:simu_sf})} that best matches the measured \SFobs~ ({the actual SF data points, not the analytic approximation obtained using Equation \ref{eq:sf}}) in the MIR through MCMC sampling with {\tt emcee} \citep{emcee}, we can test the recovery of the input torus parameters (i.e., $\overline{\sigma}$, $\overline{\theta}$, and $\overline{{\rm log}\,\Rin}$) by comparing them with that of the best-fit model SF. 

\begin{figure}
\includegraphics[width=\linewidth]{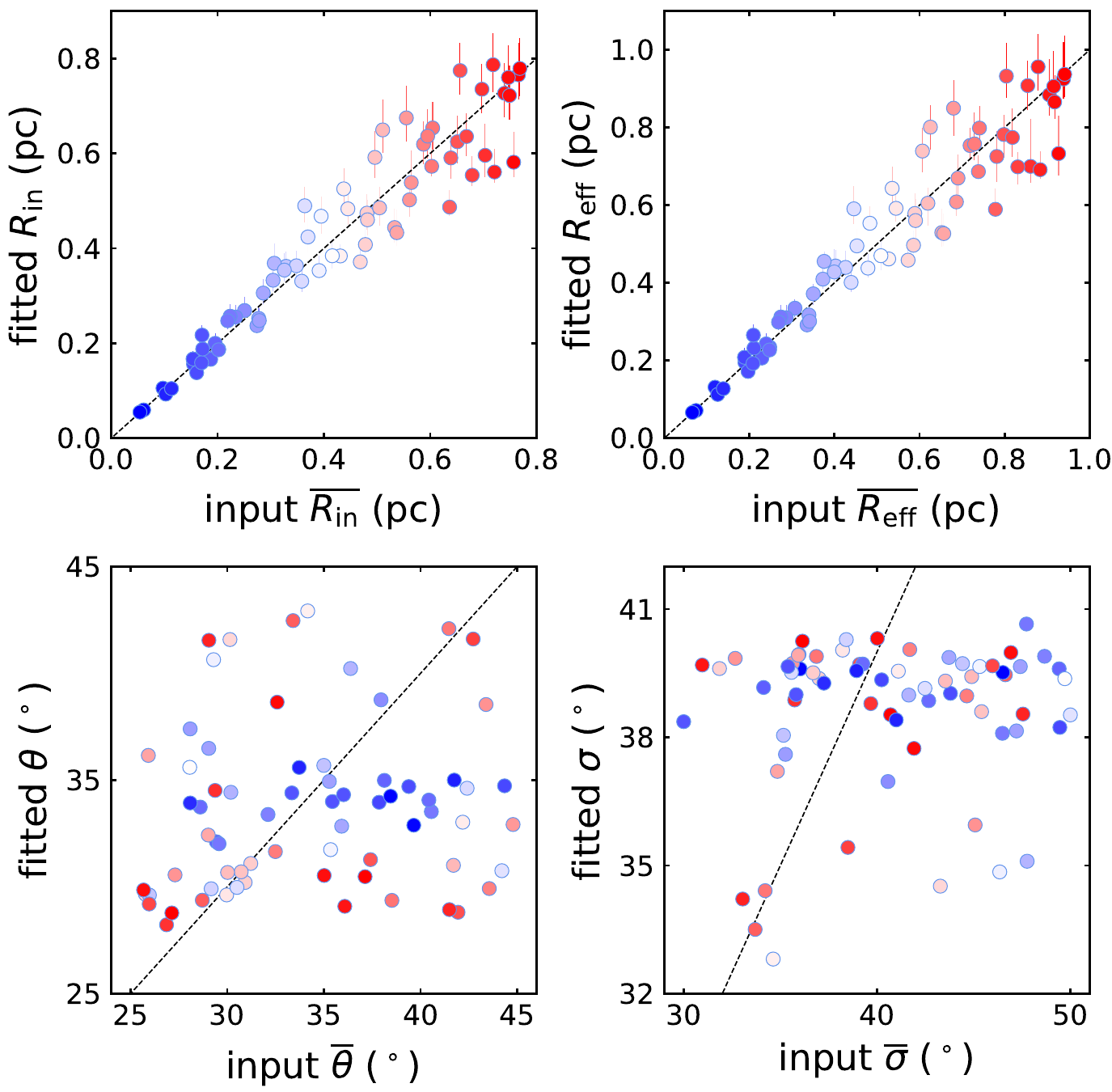}
\caption{
Recovery of the input average torus parameters for simulated light curves {with the same cadence and photometric noise as SDSS quasars. Each point represents an ensemble of 150 simulated quasars associated with a range of torus parameters with mean values of $\overline{\Rin}$, $\overline{\Reff}$, $\overline{\theta}$, and $\overline{\sigma}$, color-coded by the input $\overline{\Rin}$.} The dashed line shows the one-to-one relation. }
\label{fig:input_output}
\end{figure}

Figure \ref{fig:input_output} shows the comparison between the input and fitted torus parameters. {The best-fit parameter and its $1\,\sigma$ uncertainty are determined from the 16-50-84th percentiles of the posterior distribution (assuming a flat prior).} The inner radius of the torus can be robustly constrained, given that it has the strongest impact on shaping the SF. However, the recovery of \Rin~becomes less accurate for extended torus. This may be due to a reduction in its variability amplitude at short timescales, which makes it challenging to separate the variability signal from photometric noise. The posterior distribution of the inclination angle and opening angle spans the entire allowed fitting range, and the fitted values exhibit no correlation with their input counterparts. This failure is driven by the combined effect of the stochastic nature of SF measurement, imperfect noise subtraction, and  inherent parameter degeneracies. Nevertheless, the effective size of the torus is insensitive to the poorly constrained $\theta$ and $\sigma$, meaning that although different torus parameters are degenerate, the average response-weighted size can be well determined. 

It is noteworthy that the choice of the radius ratio $Y$ also has an impact on the derived \Rin~and $R_{\rm eff}\equiv c\tau_{\rm lag}$. If the dust emitting over a narrow range of wavelengths has a more compact (e.g., $Y=1.1$) or extended (e.g., $Y=2.0$) spatial distribution, whereas $Y$ is fixed at 1.5 in our fitting, we find that the size tends to be slightly underestimated for the former and overestimated for the latter by a factor of $\sim1.15$. Nonetheless, this small systematic difference has no significant impact on our main conclusions.

\subsection{Torus Geometry for SDSS quasars}

\label{subsec:torus}

We now apply the fitting methodology to the observed ensemble SFs of SDSS quasars to constrain their torus geometry. In the fitting, \Rin, $\sigma$ and $\theta$ are allowed to vary within the ranges of [0.01, 1.5], [30, 50], and [25, 45], as motivated by the observational constraints on the {\it average} torus parameters in Section \ref{subsec:params}. {Note that slightly changing (e.g., by 10$^\circ$) the allowed fitting range for $\sigma$ and $\theta$ has limited impact on the size measurements.} The best-fit SFs and torus parameters {in the $W1$ band} are shown in Figure~\ref{fig:sf_mag}. Our simplified torus model provides a successful fit to the data, which further support the notion that the MIR variability of quasars is driven by  UV/optical variability and diluted by the extended torus geometry. Given that the inclination and opening angle cannot be constrained with the current dataset (Section \ref{subsec:params}), we focus on the fitting result of the torus size in the following discussion.

\begin{figure}
\includegraphics[width=\linewidth]{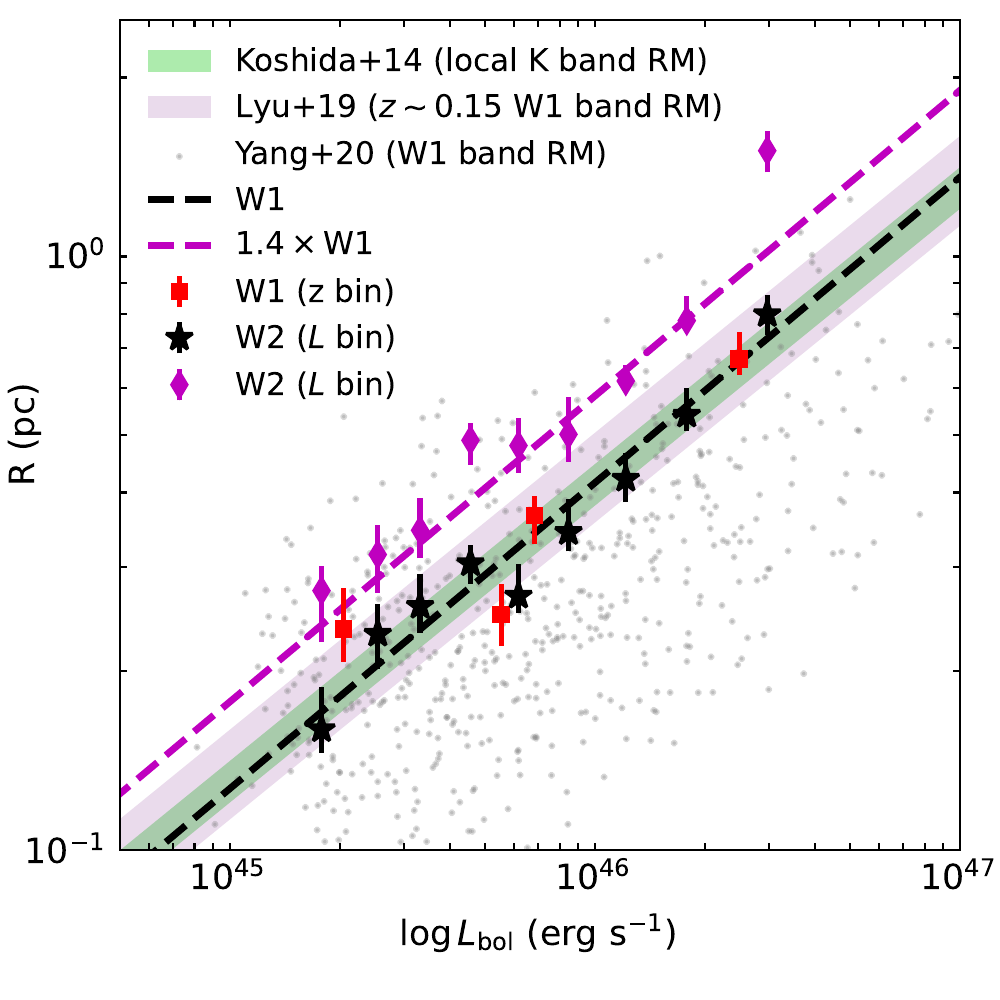}
\caption{Correlation between \Reff~and \lbol~measured from our ensemble SF method compared to literature results based on dust reverberation mapping. {The red ($W1$), black ($W1$), and magenta ($W2$) points} represent our SDSS quasars divided by redshift (i.e., S1--S4), luminosity, and luminosity, respectively, with the errorbars showing the $1\sigma$ uncertainty from the model fitting. The best-fit $R-L$ relation for the black points is shown as a black dashed line, {while the magenta dashed line represents 1.4 $\times$ black line.} The green shaded region shows the local $R-L$ relation in \cite{Koshida2014} derived from different fitting assumptions. The $R-L$ relation in \cite{Lyu2019} for different quasar populations is shown by the pink shaded region, with the upper and lower edges representing normal quasars and hot dust deficient quasars, respectively. {The \cite{Y20} sample is plotted as small gray points.} }
\label{fig:RL}
\end{figure}

In Figure \ref{fig:RL} (red points), {we present the correlation between \Reff~and \lbol~for the S1--S4 sample measured from our ensemble SF analysis in the $W1$ band.} To more effectively sample the $R-L$ relation, we further divide our quasars into nine luminosity-based subsamples (0--20th, 10--30th, ..., and 80--100th percentile of the luminosity distribution), and show them as black points. {The overlapping luminosity range is chosen to guarantee sufficient numbers of objects are included in each bin to facilitate a robust measurement of the ensemble average.} For comparison, previous results from dust reverberation mapping studies are also shown in Figure~\ref{fig:RL} \citep{Koshida2014, Lyu2019, Y20}. In particular, \cite{Lyu2019} measured time lags in the WISE $W1$ band for 67 PG quasars ($44.6 \lesssim \loglbol \lesssim 46.6$) at $\overline{z}\sim0.15$, corresponding to $\lambdar \sim 2.9\,\um$. The sample of \cite{Koshida2014} contains local low-luminosity AGNs ($42.9 \lesssim \loglbol \lesssim 45.6$) with time lag measured in the $K$ band ($\lambda_{\rm rest} \sim 2.1\um$), which is  similar to our less luminous quasars ($\loglbol < 46.0, \ \lambda_{\rm rest} \sim 1.8-2.3\,\um$) but slightly redder than the most luminous ones ($\lambdar\sim1.5\,\um$, corresponding to the $H$ band). However, simultaneous multiband monitoring of local AGNs has revealed that the time delays between the optical (e.g., $V$ band) and $JHK$ bands are similar (e.g., $\tau_{V,H}/\tau_{V,K}$ has a mean of 0.9 and a scatter of 0.13 for our compilation of $\sim20$ local AGNs (or the same AGN in different observing runs) from \citealt{Lira2015, Pozo2015, Oknyansky2015, Mandal2018, Landt2019, Sobrino2020, Mandal2021}), suggesting similar torus sizes in the rest-frame NIR bands. Therefore, our result in the $W1$ band can be directly compared to that of \cite{Koshida2014}. 

Intriguingly, the torus sizes derived from our ensemble SF method exhibit a strong correlation with luminosity. {The best-fit relation for the nine black points (neglecting correlated errors) is ${\rm log}\,\Reff\, (\rm pc) = 0.51_{-0.04}^{+0.04} \times {\rm log}\,(\lbol/10^{46}\,\ergs) -0.38_{-0.01}^{+0.01}$}. This trend is not a result of AD subtraction which steepens the SF for the most luminous quasars: a direct fit of \SFw~at $\Delta t > 200$ days, where disk variability has a lesser impact, still yields consistent result (slope $\sim 0.43_{-0.04}^{+0.04}$). 
{We also apply the same fitting methodology to the $W2$ SFs (see Appendix \ref{appendix} for details) to measure the torus sizes at a slightly longer wavelength. As shown in Figure \ref{fig:RL} (magenta points), the size in the $W2$ band is also strongly correlated with AGN luminosity, which can be well described by the $R-L$ relation in the $W1$ band but with an elevated normalization. This suggests that the $W2$ band traces dust located further away from the central SMBH. The size ratio between the two bands ($\sim1.4$) derived from our SF method is consistent with that measured from dust reverberation mapping for the same sample ($\sim1.3$; Yang et al. in prep.).}
The slope of $\sim0.5$ is consistent with the results in the low luminosity regime \citep[e.g.,][]{Koshida2014, Lyu2019} and theoretical expectation of the inner edge of the torus being determined by dust sublimation and radiation equilibrium \citep{Barvainis1987, Kishimoto2007}. Our result supports that the circumnuclear torus structures in luminous quasars and their Seyfert counterparts are shaped by the same physical process.

However, as luminosity increases, we observe a systematic difference (increase) in sizes in comparison to the torus lags measured in Y20. As discussed in Y20, the $R-L$ relation in their study may be biased low towards the high-luminosity end. This could be due to the difficulty of reliably constraining the increased time delay within a limited baseline, as well as the contamination from the accretion disk which is not subtracted in Y20. While our SF method does not suffer from this limitation, there is a substantial scatter between the input and fitted $\Reff$ for such an extended torus (Figure \ref{fig:input_output}). Therefore, we cannot rule out the possibility that our size measurements at the highest luminosities may be affected by small number statistics. To firmly measure the $R-L$ relation via this SF method, a larger sample is needed {to provide more independent ensemble SF measurements at the luminous end (e.g., a sample of $\sim1200$ quasars can double the number of independent ensemble SF measurements)}.  

\section{Concluding Remarks}
\label{sec:conclusion}

We have measured the optical and MIR ensemble SFs in the $g$ band and WISE $W1$ {and $W2$ bands} for 587 quasars at $0.2\lesssim z \lesssim2$ in the SDSS Stripe 82 region. The SFs in the MIR exhibits smaller variability amplitudes and steeper slopes than those in the optical. In addition, the MIR variability amplitude shows a negative correlation with luminosity while the slope shows a positive trend. We built a geometric torus model and demonstrated that the shape of the MIR SF and its correlation with luminosity can be well explained by the geometric dilution effect, in which the MIR variability of quasars is originated from the delayed and smoothed response of the torus to the incident UV/optical variations. We further generated simulated light curves with realistic photometric noise and cadence and demonstrated that the ensemble SFs in the MIR can be used to constrain the size of the torus. The derived size of SDSS quasars shows a remarkable  $R\propto L^{0.5}$ relation that is consistent with dust reverberation mapping measurements and theoretical expectations.

Our SF method provides an important independent means to constrain the torus $R-L$ relation, complementary to the dust reverberation mapping technique. It can be applied to light curves with any observation period, regardless of whether the optical data lead or overlap with the MIR data or not, and is less susceptible to the sampling issue, as long as the ensemble AGN variability is roughly stationary. The only requirement is to have sufficient flux pairs and precise photometric measurements to robustly measure the variability amplitude at a given timescale, which could be achieved by denser sampling, a longer baseline, or larger ensemble samples. In the next decade, the synergy between the $\sim10$ yr Vera C. Rubin Observatory and the $\sim5$ yr Nancy Grace Roman Space Telescope will further extend the baseline and provide optical and NIR light curves for AGNs up to $\sim5$~mag fainter with much higher photometric precision than the current ground-based facilities and WISE. This will not only allow more robust SF measurements for large samples of luminous quasars, but also extend the current analysis to much lower luminosities, for which the variability is intrinsically stronger \citep[e.g.,][]{VandenBerk2004} and the SFs are expected to be distinctly different (even at $\Delta t > 100 $ days) from the luminous populations given their compact torus sizes.

Finally, while the SF method is straightforward to implement and much less demanding of light curve data quality than is dust reverberation mapping, we emphasize that more geometric information describing the torus can be extracted from the full shape of the transfer function. For example, as illustrated in Figure \ref{fig:simu_sf}, although the SF is insensitive to the torus orientation given that it only measures the rms variability, the actual orientation can be constrained from the shape of the transfer function. Therefore, dust reverberation mapping using high-quality optical+IR light curves to measure the shape of the transfer function would be far more powerful to infer the full geometry of the torus. In future work, we plan to investigate the recoverability of the transfer function using optical and IR light curves, and the corresponding torus geometry with the model presented here.  

{We publicly release the code used to generate the torus models, transfer functions and SFs at this github link: \url{https://github.com/bwv1194/geometric-torus-variability}. }

\software{astropy \citep{2013A&A...558A..33A},
celerite \citep{celerite}, emcee \citep{emcee}
}

\begin{acknowledgements}
We thank the anonymous referee for a careful reading of our manuscript and for constructive comments which helped improve the quality of the paper.
This work is supported by NASA grant 80NSSC21K1566. 

This publication makes use of data products from the Wide-field Infrared Survey Explorer, which is a joint project of the University of California, Los Angeles, and the Jet Propulsion Laboratory/California Institute of Technology, funded by the National Aeronautics and Space Administration.

Funding for the SDSS and SDSS-II has been provided by the Alfred P. Sloan Foundation, the Participating Institutions, the National Science Foundation, the U.S. Department of Energy, the National Aeronautics and Space Administration, the Japanese Monbukagakusho, the Max Planck Society, and the Higher Education Funding Council for England. The SDSS Web Site is http://www.sdss.org/. 
\end{acknowledgements}

\bibliography{sample631.bbl}{}
\bibliographystyle{aasjournal}

\appendix 
\section{Structure Function in the $W2$ band and the impact of imperfect noise subtraction}
\label{appendix}

\begin{figure}
\includegraphics[width=\linewidth]{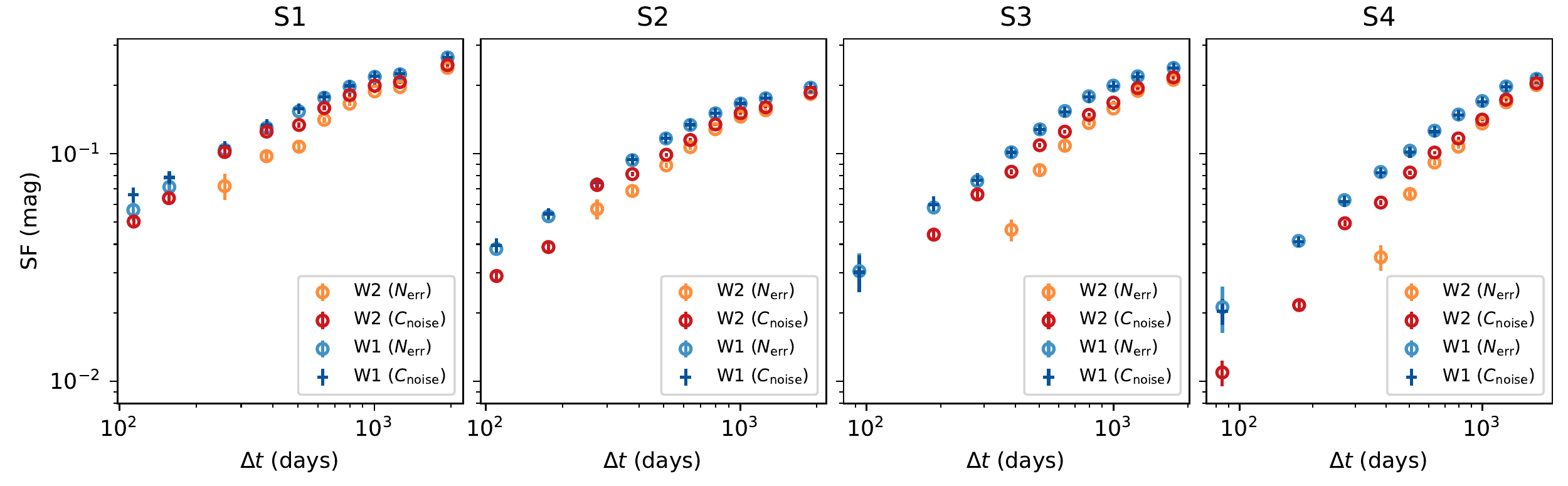}
\caption{Ensemble SFs in the $W2$ band compared to that in the $W1$ band measured from different noise subtraction methods for the S1--S4 samples. }
\label{fig:w2_sf}
\end{figure}

\begin{figure}
\centering
\includegraphics[width=0.5\linewidth]{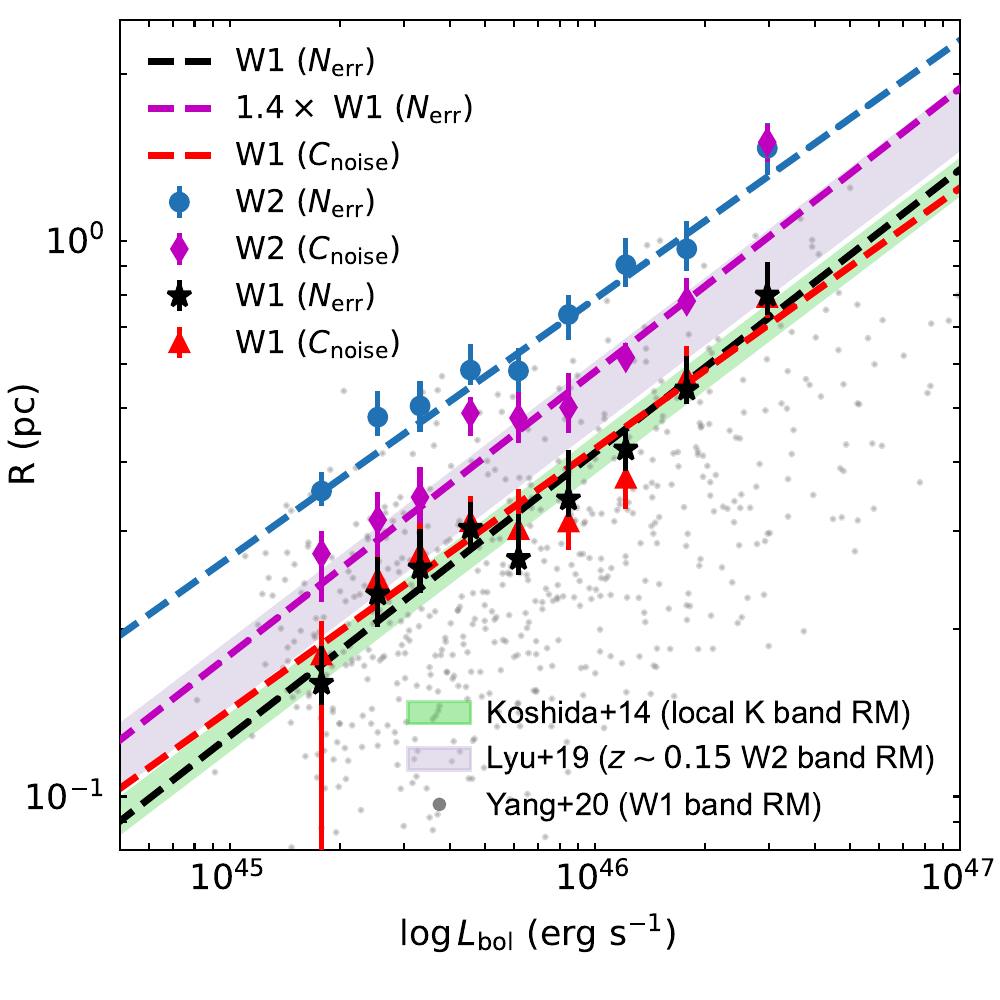}
\caption{The $R-L$ relation in the $W1$ and $W2$ bands measured from different noise subtraction methods compared to literature results (similar to Figure \ref{fig:RL}).}
\label{fig:RL_w2}
\end{figure}

\begin{figure}
\centering
\includegraphics[width=0.8\linewidth]{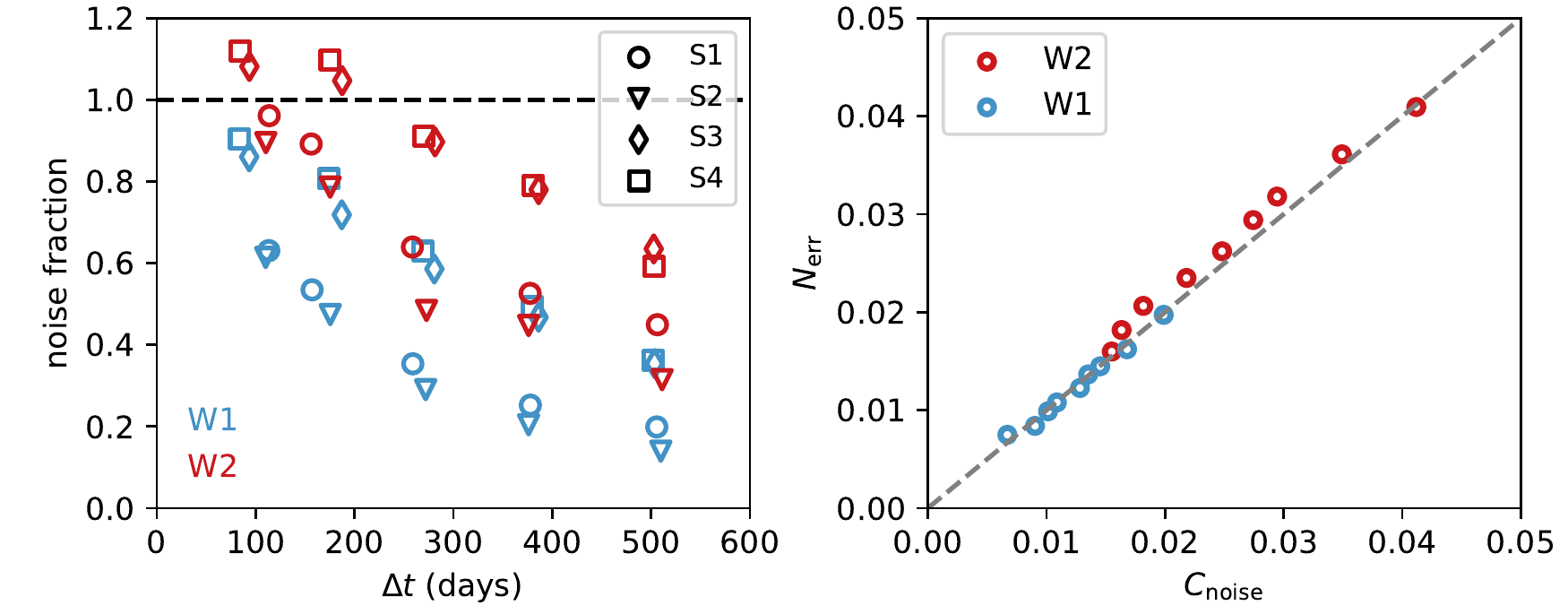}
\caption{Left: noise fraction (i.e., \nerr/observed rms variability) at different timescales in the $W1$ (blue) and $W2$ (red) bands. The S1--S4 samples are indicated by different markers. Right: comparison between the noise term estimated from the observed flux uncertainties (\nerr) and that derived from the functional fitting method (\cerr).}
\label{fig:w2_noise}
\end{figure}

This appendix presents the details of SF measurements in the $W2$ band compared with the $W1$ results shown in the main text. Additionally, we discuss the impact of imperfect noise subtraction on the SF measurements and the derived torus sizes. Figure \ref{fig:w2_sf} displays the ensemble SFs (in magnitude units) in the $W2$ band measured from the \nerr~method (i.e., noise term estimated from the reported photometric uncertainties) and compare them with those in the $W1$ band. As expected, the $W2$ SFs are steeper and have smaller variability amplitude, given that the $W2$ band traces a colder and more extended torus component. Additionally, the $W2$ slope exhibits a similar positive correlation with AGN luminosity. However, the $W2$ SF appears to be excessively steep, even becoming negative at $\Delta t \lesssim 300$~days. Applying the fitting methodology in Section \ref{subsec:torus} to the $W2$ SFs (measured in flux units and excluding negative points) results in a strong $R-L$ relation that is similar to the $W1$ band (blue points in Figure \ref{fig:RL_w2}). But the derived $W2$ size is $\sim2$ times larger than the $W1$ size, which significantly exceeds that measured from dust reverberation mapping for the same sample ($\sim 1.3-1.4$; Yang et al. in prep.).

The steep/negative SF in the $W2$ band is likely caused by its elevated flux uncertainty ($\sim1.4$ times larger than the $W1$ band). Ideally, this elevated photometric uncertainty will cause an increase in both the observed rms variability and \nerr, while the subtraction of the latter from the former will still return an unbiased estimate of the intrinsic variability, which should be strictly greater than 0. However, our analysis reveals that the noise fraction (i.e., \nerr/observed rms) in the $W2$ band can be even greater than 1.0 (left panel in Figure \ref{fig:w2_noise}), indicating that the photometric uncertainties are overestimated.

The comparison between \nerr~to the noise term derived from the \cerr~method based on functional fitting (Section \ref{sec:ensemble}) in the $W2$ band is shown in Figure \ref{fig:w2_noise} (right panel). In order to obtain a more accurate estimation of \cerr, we imposed a requirement in the fitting  that the $W2$ slope $\beta$ should be steeper than the $W1$ slope, which is more accurately constrained. The fitted \cerr~is $\sim1.07$ times smaller than \nerr. As a result, the $W2$ SFs measured by subtracting \cerr~(red points in Figure \ref{fig:w2_sf}) become only slightly steeper than the $W1$ SFs, and negative values are largely avoided. Consequently, the $W2$ torus sizes are reduced by a factor of $\sim1.4$ (magenta points in Figure \ref{fig:RL_w2}). This reduction brings the $W2/W1$ torus size ratio ($\sim1.4$) into agreement with that determined from reverberation mapping. Our analysis highlights the crucial role of accurately estimating the noise term for the measurement of SF and the determination of torus size. Even a subtle overestimation of the photometric noise (in this case, only $\sim 7\%$ overestimation) in the $W2$ band can substantially affect the inferred SFs, as the noise contribution dominates the observed rms variability at these shortest timescales (Figure \ref{fig:w2_noise}). The empirical \cerr\ method properly removes the constant SF floor at short timescales due to photometric errors, and does not depend on the accuracy of photometric errors in $W2$.

In contrast, the two noise estimation methods yield consistent results in the $W1$ band (Figure \ref{fig:w2_noise}), leading to almost identical SFs (dark and light blue points in Figure \ref{fig:w2_sf}) and $R-L$ relations (red and black lines in Figure \ref{fig:RL_w2}). This agreement suggests that the reported photometric uncertainties in the $W1$ band are much more accurate than those in $W2$.

\end{document}